# Validation of Collaborative Business Processes using Goals Model

Amir Ebrahimifard
PhD Student, Department of Technology, Policy and Management, TU Delft University,

Mostafa Khoramabadi Arani
MS Student, Department of Computer Engineering, Iran University of Science and Technology,

Mohammad Javad Amiri
PhD Student, Department of Computer Science, University of California Santa Barbara, USA,

Saeed Parsa
Professor, Department of Computer Engineering, Iran University of Science and Technology,

*Abstract*—Validating process model against corresponding requirements is one of the most important problems in domain of collaborative processes. In this paper collaborative processes are modeled using the interaction view of BPMN 2.0 standard. Then, requirements are extracted with a goal modeling technique. Different scenarios of each requirement show possible paths for the system. These paths are modeled by sequence diagram and collaborative processes are validated according to the corresponding requirements using Savara tool.

## 1. Introduction

A collaborative process shows a collaboration between different processes to provide a composite service. Service composition is the process of developing a composite service from atomic services [1]. The aim of web service composition is to arrange multiple services into collaborative workflow supplying complex user needs [2]. Today, with service composition, instead of developing monolithic applications, we build large-scale software applications by composing loosely coupled services [3]. By doing so, we can reuse and select suitable services as various organizations can provide similar services for the development of different applications [4]. There are two possible ways for service composition: orchestration and choreography that are defined as two aspects of inter- and intra-organization service composition.

Choreography describes a system in a top view manner in which multiple cooperating independent agents exchange messages in order to perform a task to achieve a goal state whereas orchestration focuses on single peer's description [5]. Choreography basically uses to represent a collaborative business process.

Choreographies are particularly useful in those situations in which multiple parties have to collaborate, but none of them wants to take the responsibility of running a centralized orchestration [6].

One of the most important issues in choreography service composition is choreography validation. It is defined as accordance between requirements and choreography definition. In other words, to what extent a choreography definition can explain and justify requirements specifies validity level of that choreography definition [7]. Currently in the context of choreography validation, lack of a comprehensive validation method to validate model against the requirements is counted as a major issue.

To overcome the abovementioned issue, requirements are considered as the base of choreography model validation in the proposed method. To this end, goals modeling technique is used to extract the requirements of the target process. Then, for each requirement all possible scenarios are described. These scenarios are modeled using Savara tool. In the next step the choreography process is modeled using BPMN 2.0 interaction modeling. By modeling the process, the next step is extracting all possible paths from the model and matching each path with the corresponding requirement. If all requirements are satisfied, that is all possible scenarios occurred and there are no extra paths in the model, the model is valid.

This paper is organized as follows. Requirement extraction using goal modelling technique is presented in Section 2. The focus of Section 3 is on process choreography modeling. Section 4 provides information to validate a choreography model. In Section 5 an evaluation of our method is presented and conclusions are included in Section 6.

## 2. Requirement Extraction

In this section a goal modelling technique is introduced to extract requirements. To start with, we introduce a Purchasing process which is used as a running example in this paper.

**Running Example:** In a purchase process three parties are participated; buyer, agency, and factory.
A buyer corporation sends a registration request. If the type of request is ordinary, the request does not need to be confirmed, but if there is a special request, the registration should be confirmed by the agency. Also, if the type of request is superior, the agency sends the request to the factory and the factory gets the information from the agency and confirms the request. In all three cases, the buyer should start the registration sub-process. After the registration, the buyer can order the request to factory. The factory should confirm this request and until the confirmation of the request, data transfer continues between the factory and the buyer. After request confirmation, the company informs the buyer and the buyer in contact with the agency, can determine the delivery type. Then the factory informs the buyer about the cost and the buyer pays this price. If the payment is not successful, the factory informs the buyer and sends a no deliver request to agency.

Goals model represents goals, objectives and requirements of the system. A goal is a purpose to be achieved by the system under consideration [8]. Goals modeling is one of the requirements elicitation techniques that help the system analyst to recognize all requirements to be fulfilled by the software product. Goal models have been used to model requirements for software systems, business objectives and design qualities [9]. In a goals model, high-level goals are decomposed to sub-goals until requirements are extracted. Creating the goals model is a difficult task and needs the cooperation of the representatives of developers, business workers and even users [10].
In the first step of the proposed method, goals modeling technique is used to extract requirements. To do this, the goals model is created in a simple way and goals are shown without any property.
We define a goal model as a set of goals where one of them is the final goal of a system. Requirements are the leaves of the goal models and there is a set of edges between goals that show the goal refinement.

**Definition:** A Goal Model is tree denoted by the tuple *M = (G, f, R, E)* such that
- *G* is a non-empty set of goals,
- *f ∈ G* is the final goal which is root of the tree,
- *R ⊆ G* is the set of requirement which are leaves of the tree, and
- *E ⊆ (G-R) × (G-f)* is a finite set of edges.

**Example 2.1.** *Figure 1* shows a simple goals model *M = (G, f, R, E)* for a purchasing process.
In this goals model there are totally nine goals in G where *f = Purchasing* is the final goal. *Purchasing* has three sub goals, *Customer Authentication, Order Processing*, and *Financial Processing*. *R = {Registration, Order Delivery, Providing Order, Price Negotiation, Payment}* is the set of requirements and finally there are some edges between goals such as *(Purchasing, Order Processing)* and *(Financial Processing, Payment)*.

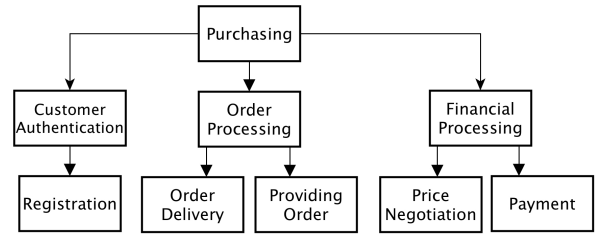
Figure 1: Purchasing Goals Model

By extracting the requirements of process, the scenario of each requirement can be specified. Each requirement has different scenarios that show all possible ways to satisfy that requirements. From the choreography point of view, a scenario should contain the required collaboration between different participants based on different situations.

**Example 2.2.** Continuing with *example 2.1.* for the registration requirement, four different scenarios can be described.
There are four possible scenarios for the registration requirement.
1) Ordinary Registration: The buyer sends an ordinary registration request to the agency and then enters the registration process.
2) Special Registration: The buyer sends a special registration request to the agency. The agency should confirm the request and then the buyer enters the registration process.
3) Superior Registration: The buyer sends a superior registration request to the agency. The agency in iterative contact with the factory confirms the request. After confirmation, the agency informs the buyer and the buyer enters the registration process.
4) Invalid request: The buyer sends a buying request to the factory. The request is invalid, so the buyer asks the shortcoming list from the factory and resends the buying request. This cycle continues until the request is confirmed. Then the documents should be sent to the factory.
These scenarios can be modeled using sequence diagram where each participant models as an object and collaborations between participants are shown.

**Example 2.3.** Continuing with the *example 2.1.* A sequence diagram for successful credit check scenario can be seen in *Figure 2*. Savara tool is used to model these scenarios.

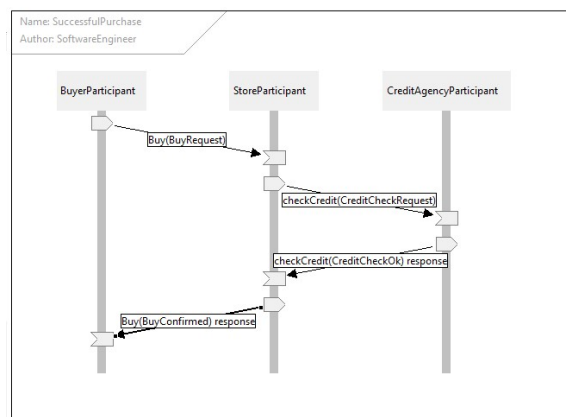
Figure 2: Successful Credit Check Scenario

## 3. Process Choreography Modeling

In this section business process choreography, which shows the cooperation of participants, is modeled. A business process is defined as a set of tasks performed in coordination to achieve business objectives [11]. Also choreography and orchestration describe two aspects of web service composition [12].

Orchestration is an executable business process built with Web Services seen from a single-party perspective, while choreography is the message sequence between multiple web services seen from the perspective of multiple parties [13]. The relation between these two concepts is that the interaction between individual behaviors of each involved party, which is defined via the orchestration, results in a collaborative behavior, described by the choreography [14]. W3C defines choreography as the sequence and conditions under which multiple cooperating independent agents exchange messages in order to perform a task to achieve a goal state [15].

There are two main approaches for choreography modeling: inter-connection modeling and interaction modeling. In inter-connection modeling, data and dependencies are defined within each role, and message sending and reception are in separate activities, whereas in interaction modeling, atomic message exchanges are the basic building blocks and control and data flow are defined globally [16].

Using interaction modeling results in several advantages for accurate choreography modeling:

(a) Control flow dependencies are not defined per role, but rather seen from a global perspective. In this way, redundancy in control flow relationships is avoided and the interactions ordering prevents modeling anti-patterns and deadlock. Having less redundancy in structures leads to faster and easier modeling.

(b) Global specification of branching structures avoids modeling errors such as decision making anti-patterns, and process instance creation and termination anti-patterns caused by incompatible branching structures [17].

There are some challenges that are caused by interaction modeling which did not exist before, because dependencies no longer belong to individual partners and are defined on a global level.

In this paper due to the closeness to the business goals and requirement, the use of business process modeling notation (BPMN 2.0[18]) is suggested. Simplicity, expressiveness, and high usage have been the main advantages of this standard, compared with other methods of business process modeling [19]. Also [20] provides a complete study on process choreography languages and shows the suitability of the BPMN 2.0 for choreography modeling. Let $M$ be an infinite set of all names. A choreography model can be defined as follow [21].

**Definition:** A choreography model is a directed, acyclic graph denoted by the triple *(m, P, L),* where
- $m \in M$ is the name of the choreography model,
- $P$ is the set of choreography participants, and
- $L$ is the set of message links between the choreography participants.

**Definition:** A message link in $L$ is a tuple *l = (t, $p_s$, $p_r$, $a_s$, $a_r$)* where
- $t \in M$ is the message name,
- $p_s, p_r \in P$ are the sending and receiving participants where $ps \neq pr$, and
- $a_s \in A_s$ and $a_r \in A_r$ are the sending and receiving activities in $p_s$ and $p_r$ respectively where $A_s$ and $A_r$ are the sets of activities in $p_s$ and $p_r$.

basically each message link has a sender and a receiver which are choreography participants. Each message has a corresponding send activity in the sender party and a corresponding receive activity is the receiver party.

**Example 3.1.** *Figure 3* shows a BPMN 2.0 choreography model for the abovementioned purchasing process.

## 4. Choreography Model Validation

In this section, a method for validating the choreography model is introduced. Validation is defined as determination of correctness of the final program or software produced from a development project with respect to the user needs and requirements [22].

In the analysis of web service specification, an important issue is checking that the given composition of existing services satisfies the requirements of the specified global choreography protocol [23]. In other words, an important problem in web service domain is choreography validation. The importance of the validation in choreography is the same as the importance of validation in software development process. In software development process, most of the programmers' effort is lost in extensive and often repeated testing lifecycles because of ambiguity in capturing and analysis of requirements and then ambiguity between architecture and requirements and finally the cascading effect of ambiguity between implementation and architecture.

Empirical research has shown that the cost of correcting design defects at the traditional testing stage is around 200 percent more than correcting them during the requirements or architecture stage. [24]. Therefore, if choreography model can be validated against requirements, the costs of software testing will drop dramatically. In this work, the main goal is to know whether BPMN 2.0 choreography model fully implements requirements or not. By knowing this, choreography designer makes progress confidently, and the probability of changes in basic part of choreography (like designing global model) becomes a small number.

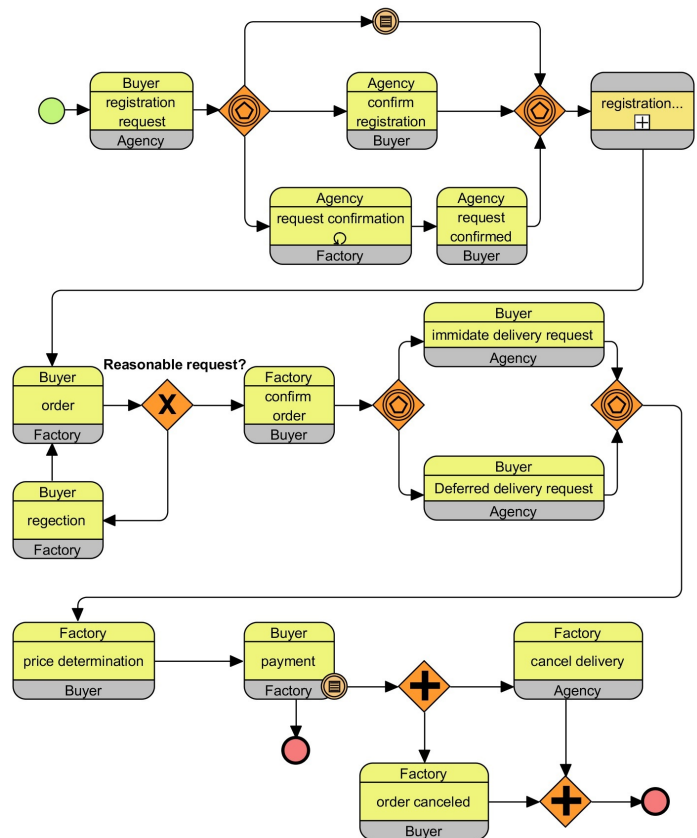

Figure 3: Purchasing Process Model

### 4.1. Choreography Validation Requirements and approaches

As discussed in [23] choreography validation has two main requirements:

1. Conformance checking: it has to check that the behaviors exhibited by the composition correspond to those described in the choreography document.

2. Information alignment: it is often needed that the participants agree on the state of the protocol as a result of its execution. In other words, they expect to have a common knowledge on certain variables that describe the state of the protocol.

To satisfy these two requirements, choreography validation approaches are introduced. As discussed in [25] choreography validation has two groups of approaches: bottom-up and top-down. In bottom-up approaches, it is important to know whether individual peers implemented as services and advertising behavioral descriptions (conversations) are behave as describes in the choreography or not. Knowing this issue is critical for continuing development process. Like bottom-up approach, the validation in a top-down development process is also central.

In recent decade researchers suggest various solutions for choreography validation. One of the best solutions is testable architecture. Testable Architecture was developed by Red Hat and Cognizant Technology Solutions. As mentioned earlier, there is ambiguity because requirements are not satisfied with architecture and architecture is not satisfied with implementation. Testable architecture is the foundation of removing this ambiguity. It enables the architecture of a system to be described unambiguously using Choreography Description Language (CDL) such that it may be tested against requirements. Testable architecture is formally grounded with strong type definition and has its foundations in pi-calculus [24]. It aims to ensure that any artifacts defined during the development lifecycle can be validated against other artifacts in preceding and subsequent phases. Also this methodology ensures that the final delivered system is guaranteed to meet the original business requirements [26].

### 4.2. The Savara Tool

The Savara project has been established to create tool support for Testable Architecture [24]. This project aims to leverage the concept of a choreography description to provide design-time and run-time governance of an SOA [27]. Savara aims to dramatically reduce testing expenditure and overall software development costs through modeling and simulation and makes it enterprise scale. Savara ensures that artifacts defined in each phase of the software development lifecycle can be verified for conformance. It means architectural models can be verified against requirements, service designs against architectural models and code against service designs. This guarantees that the deployed systems can be shown to implement the originating business requirements [24]. Savara provides tools for designing choreographies using BPMN 2.0 notation and deriving partner business processes (expressed in BPEL) [28].

### 4.3. Model Validation

After modeling requirements and choreography, validating phase between requirements and choreography can be done that means the Global Model can be checked for conformance against requirement Scenarios. At this phase, namely conformance phase, due to the previously defined global model of choreography, conformance between requirements and global model of choreography is applied. In this phase correct and incorrect paths that are specified in pervious steps, are checked against the choreography model. A correct choreography model must implement correct paths and must not implement incorrect paths. If the opposite of each case occurs, then choreography model has problems and must be modified.

After these steps, level of accordance between requirements and global model is specified and if there is a requirement that choreography model doesn't support it, then the model must be corrected.

**Example 4.1.** Figure 4 shows the correct scenario for unsuccessful payment that is modeled from when the factory announced the cost to the buyer.

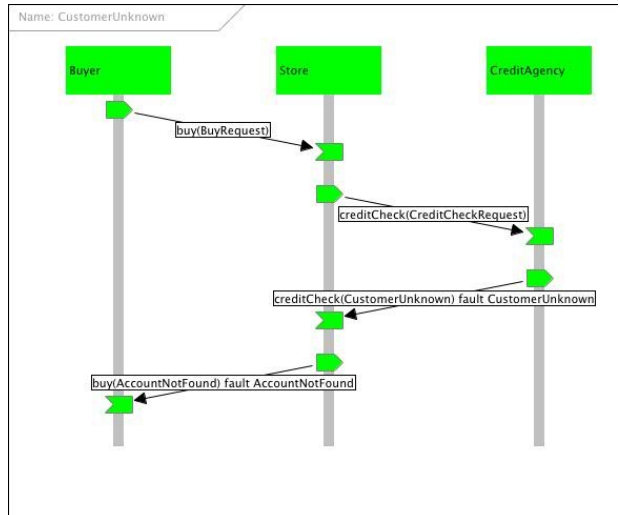
Figure 4: Correct Scenario of Unsuccessful Payment

**Example 4.2.** Figure 5 shows the incorrect scenario for order delivery that cannot be performed, because the customer is unknown.

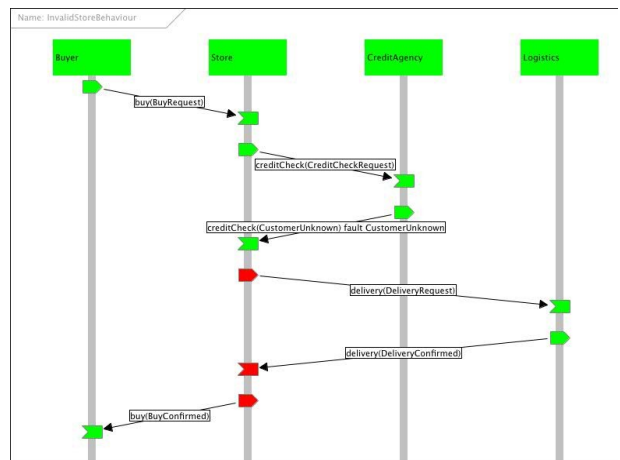
Figure 5: Incorrect Scenario of Order Delivery

## 5. Results and Discussion

In order to evaluate the proposed method, several processes description and their equivalent choreography models have been selected from MIT Process Handbook [29] in various domains. Then in two separate steps these models are given to our proposed method and some modelers for validation. Modelers tried to identify the choreography model and requirements to what extent are consistent with each other. To measure the truth of their assessment, four different metrics have been considered:
1. Valid paths that are correctly detected as valid (True Positive or TP)
2. Valid paths that are incorrectly detected as invalid (False Negative or FN)
3. Invalid paths that are incorrectly detected as valid (False Positive or FP)
4. Invalid paths that are correctly detected as invalid True Negative or TN)

The results of users' evaluation have brought in the Table 1.

I. Validation by users

| Process | Process Specifications | | | By Users | | | |
|---|---|---|---|---|---|---|---|
| | # of requirement | # of valid path | # of invalid path | True Positive | False Negative | False Positive | True Negative |
| Replenish inputs | 3 | 12 | 1 | 9 | 3 | 0 | 1 |
| Create new market space | 5 | 21 | 5 | 17 | 4 | 3 | 2 |
| Optimize the supply chain | 6 | 16 | 2 | 10 | 6 | 2 | 0 |
| Manage raw material inventory | 3 | 7 | 0 | 5 | 2 | 0 | 0 |
| Recycle and manage product returns | 4 | 11 | 5 | 6 | 5 | 3 | 2 |
| Forecast demand with suppliers | 4 | 9 | 4 | 4 | 5 | 1 | 3 |
| Collect sales data at POS | 2 | 5 | 3 | 3 | 2 | 0 | 3 |
| Hire human resources | 6 | 18 | 12 | 16 | 2 | 4 | 8 |
| Pay employee | 3 | 8 | 5 | 6 | 2 | 2 | 3 |
| Manage risk by outsourcing | 4 | 11 | 3 | 7 | 4 | 3 | 0 |

In order to evaluate the reliability of users' measurements, precision, recall and accuracy are then defined as:

$$Precision = \frac{TP}{TP + FP}$$
$$Recall = \frac{TP}{TP + FN}$$
$$Accuracy = \frac{TP + TN}{TP + TN + FP + FN}$$

These measures have been calculated for each process and the results can be seen in Table 2.

II. Accuracy, Precision and Recall for each process

| Process | Accuracy | Precision | Recall |
|---|---|---|---|
| Replenish inputs | 77% | 100% | 75% |
| Create new market space | 73% | 85% | 81% |
| Optimize the supply chain | 56% | 83% | 63% |
| Manage raw material inventory | 71% | 100% | 71% |
| Recycle and manage product returns | 50% | 67% | 55% |
| Forecast demand with suppliers | 54% | 80% | 44% |
| Collect sales data at POS | 75% | 100% | 60% |
| Hire human resources | 80% | 80% | 89% |
| Pay employee | 69% | 75% | 75% |
| Manage risk by outsourcing | 50% | 70% | 64% |

The average amount of Accuracy, Precision and Recall for all processes are as follows: 66%, 84% and 68%. Given that the proposed method can detect all the paths correctly, we are witnessing 34%, 16% and 32% progress for Accuracy, Precision and Recall of validation in our method. The only drawback of the proposed method is that the process of requirements modeling is time consuming.

## 6. Conclusion

In this paper a method to validate web service choreography model is presented. To this end after extracting the related requirements using goals modeling technique, the scenarios of these requirements are explained and these scenarios are modeled using sequence diagrams. By modeling the web service choreography, the model is validated against the requirements' scenarios using Savara tool. Our experiments show that users make a lot of mistakes in validation process, so the proposed method can help them a lot. As a possible future work we want to extend the validation technique by taking data into account. There have been various advantages of modelling data dependencies in processes which are discussed in [29]. We may be interested in checking whether a process enactment uses correct data to achieve a goal or not.